# Energy Management of Airport Service Electric Vehicles to Match Renewable Generation through Rollout Approach

Renjie Wei [a], Kang Ma [a]

[a]*Department of Electronic & Electrical Engineering, University of Bath, Bath BA2 7AY, UK*

*Correspondence author: Kang Ma. Email: K.Ma@bath.ac.uk    Tel: +44(0)1225385463*

***Abstract***: Traditional diesel-based airport service vehicles are characterized by a heavy-duty, high-usage-frequency nature and a high carbon intensity per vehicle per hour. Transforming these vehicles into electric vehicles would reduce $CO_2$ emissions and potentially save energy costs in the context of rising fuel prices, if a proper energy management of airport service electric vehicles (ASEVs) is performed. To perform such an energy management, this paper proposes a new customized rollout approach, as a near-optimal control method for a new ASEV dynamics model, which models the ASEV states, their transitions over time, and how control decisions affect them. The rollout approach yields a near-optimal control strategy for the ASEVs to transport luggage and to charge batteries, with the objective to minimize the operation cost, which incentivizes the charging of the ASEVs to match renewable generation. Case studies demonstrate that the rollout approach effectively overcomes the "curse of dimensionality". On both typical summer and winter days, the rollout algorithm results in a total cost approximately 10% less than that of the underlying "greedy charging" heuristic, which charges a battery whenever its state of charge is not the maximum. The rollout algorithm is proven to be adaptive towards flight schedule changes at short notice.

# 1. Introduction

The transition to electric vehicles (EVs) is vital for fulfilling the target of reducing $CO_2$ emissions by 80% by 2050 in the UK, relative to the 1990's level [1]. Much attention was devoted to electrifying tens of millions of consumer vehicles. Although they are vast in number, they have a relatively low carbon intensity in terms of emission per vehicle per hour, because an average consumer vehicle remains dormant in most hours of a day and it is of a light-duty nature. Unlike consumer vehicles, airport service vehicles are characterized by a heavy-duty, high-usage-frequency nature, a high carbon intensity per vehicle per hour, and a strong correlation with flight patterns. Transforming diesel-based airport service vehicle fleets into EVs would dramatically reduce $CO_2$ emissions for this carbon-intensive industry. In this context, airport service electric vehicles (ASEVs) specifically refer to electric trailers that transport checked luggage between the sorting facility in the terminal and departure/arrival flights. The aim of this paper is to develop an optimal energy management strategy for the ASEVs in terms of battery charging and task assignment.

Existing research work focuses on consumer EVs and taxis at different locations, e.g. households, office buildings, highway service stations, etc. References [2], [3] focus on consumer EV charging at commercial buildings. A number of references consider domestic EVs as a part of home energy management systems [4], [5], [6], an energy hub [7] or a community energy system [8]. A number of references investigate the operation of electric vehicle parking lots [9], [10], including airport parking lots [11]. References [12], [13] both develop stochastic optimization models for the joint operation of EV fleets and renewable generation. Reference [14] develops a balanced charging strategy to satisfy both the EV owners (saving costs) and the network operator (relieving loads). Reference [15] develops an optimization model to schedule airport ground operations, including aircraft and shuttle bus scheduling. Although that reference does not focus on EVs, it acknowledges the importance of the optimization of airport ground operations.

Because consumer EVs are owned by many different individuals, their behavior reflects human lifestyles as well as individual random behavior. However, ASEVs demonstrate fundamentally different behavior as compared to consumer EVs, because ASEVs are centrally controlled and their behavior depends on the flight schedule, the number of passengers, and luggage weight. Therefore, ASEVs constitute a dynamic system of a stochastic, dynamic, hybrid nature that is distinct from consumer EVs and not reported in existing literature. The uncertainty in the ground transport workload renders the model a stochastic nature. The existence of both discrete variables (the decision variables for individual ASEVs to undertake ground transport tasks, charge, or idle) and continuous variables (the battery state of charge) renders the model a hybrid nature. Both types of variables change over time, rendering the model a dynamic nature. It should be noted that, although the ASEV dynamics shares a similar stochastic, hybrid, and dynamic nature with a stochastic hybrid system (SHS) [16], it is not a SHS, because the discrete variables of the ASEV dynamics system do not follow a controlled Markov chain.

There are ASEV suppliers [17], [18], [19], but the optimal control of the ASEVs was an unanswered question. For an airport with tens of ASEVs, the dynamic system has a prohibitively large number of states (i.e. the "curse of dimensionality"), too large to derive an accurate optimal solution to the ASEV control problem. Therefore, two research questions arise from the optimal scheduling of ASEVs: 1) the modeling of the ASEVs as a distinctive dynamic system of a stochastic dynamic, hybrid nature; and 2) the derivation of an optimal control strategy for the dynamic system.

The optimal control of a dynamic system is related to stochastic dynamic programming [20] in terms of their stochastic and dynamic nature. The rollout algorithm for dynamic programming [20], [21] can be borrowed but it needs to be adapted for the optimal control of ASEVs: the underlying heuristic control strategies need to be defined and uncertainties need to be properly modeled.

In summary, the following gaps are identified from the literature survey:

1) The optimal management of ASEVs is a problem not investigated before.

2) There is an absence of an ASEV dynamics model, which describes the ASEV states, their transitions over time, and how control decisions affect them.

3) There is an absence of an energy management method, which controls the ASEVs to meet a low-cost, low-carbon objective, subject to the ASEV dynamics.

To bridge the above gaps, this paper makes the following original contributions:

1) This paper proposes a new ASEV dynamics model. The model involves: i) discrete dynamics, i.e. the changes of the ASEV discrete states to "work", "charge", or "idle" over time; ii) continuous dynamics, i.e. the changes of the battery state of charge (SoC) over time; and iii) a stochastic nature of the ground transport workload.

2) To perform an energy management of the ASEVs, this paper proposes a new customized rollout approach, as a near-optimal control method for the ASEV dynamics model. The approach controls the ASEVs to transport luggage and to charge batteries, with the objective to minimize the total operation cost. Two customized suboptimal heuristic control strategies are proposed as the base strategies for the rollout approach, which then iteratively improves the heuristic control strategies into a near-optimal control strategy. The rollout approach effectively overcomes the "curse of dimensionality" challenge.

The energy management of ASEVs through the rollout approach will bring a number of benefits: 1) it will save costs for the airport; 2) by matching the ASEV battery charging load curve with renewable generation, the control method encourages the ASEVs to consume locally generated renewable energy, reduces $CO_2$ emissions, and makes the charging load curve friendly to the grid.

The rest of this paper is organized as follows: Section 2 gives an overview of the methodology; Section 3 presents the ASEV dynamics model; Section 4 presents the optimal control method for the ASEV dynamics model; Section 5 performs case studies; and Section 6 concludes the paper.

## 2. Overview of methodology

The optimal control of ASEVs aims to minimize the total operation cost, including the energy cost and battery degradation cost. Electrical energy comes from two sources: 1) energy from the grid under time-based tariffs; and 2) energy purchased directly from local renewable generation. The 2nd energy source has a lower tariff than the 1st source. This encourages ASEVs to consume locally generated energy for local balancing.

The ground transport workload depends on the number of passengers and luggage weight for each flight. The uncertainties in the ground transport workload are incorporated into the ASEV dynamics model, which models the control decisions, the ASEV states and their transitions.

As a near-optimal control method for the ASEV dynamics model, the rollout approach starts from two suboptimal heuristic control strategies and iteratively improves the better one of the two heuristics towards reducing the total operation cost.

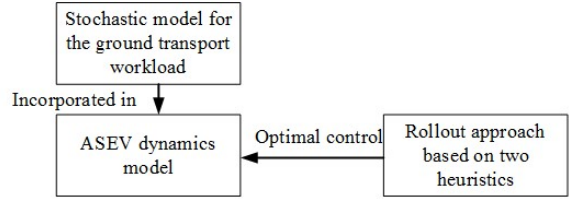

Fig. 1. Flowchart of the models and the optimal control approach

Fig. 1 shows a flowchart consisting of the three components: the stochastic model of the ground transport workload, the ASEV dynamics model, and the rollout approach.

## 3. Problem formulation: ASEV dynamics model

### 3.1 Modeling of Uncertain Ground Transport Workload

Before presenting the ASEV dynamics model, the ground transport workload model is presented. Suppose the $j$th flight is awaiting ground transport service at time $t$ (called flight $j$ at time $t$), because it has landed or is ready to depart. The time required for an ASEV to serve this flight is stochastic because: 1) although the airline company knows the number of passengers and luggage weight for the flight in question, the information may not be shared with the airport. 2) Even if the information were made available to the airport, there is a random noise in the time required to service the flight. Denote the time required for an ASEV to serve flight $j$ at time $t$ as $\widetilde{w}_{jt}$, which follows a truncated normal distribution $\psi$ [22].

$$\psi(\mu, \sigma, a, b, \widetilde{w}_{jt}) = \begin{cases} 0 & \text{if } \widetilde{w}_{jt} \leq a \\ \dfrac{\emptyset(\mu, \sigma^2; \widetilde{w}_{jt})}{\Phi(\mu, \sigma^2; b) - \Phi(\mu, \sigma^2; a)} & \text{if } b < \widetilde{w}_{jt} < a \\ 0 & \text{if } \widetilde{w}_{jt} \geq b \end{cases} \quad (1)$$

where $\mu$ and $\sigma$ are the mean and deviation of the "parent" normal distribution, respectively. a and b are the upper and lower bounds, respectively. $\emptyset(\mu, \sigma^2; x)$ and $\Phi(\mu, \sigma^2; x)$ are the probability density function and cumulative distribution function, respectively, of the "parent" normal distribution with mean $\mu$ and deviation $\sigma$. The truncated normal distribution model is justified because: 1) a normal distribution is a default choice when there is no detailed knowledge to support alternative complicated probability distributions; and 2) $\widetilde{w}_{jt}$ is bounded in reality.

Suppose that the 24 hours of a day are discretized into 288 stages, starting from Stage 0 to Stage 287 at an interval of 5 minutes. Let $w_{jt}$ denote the discrete number of stages (essentially the amount of time) required for an ASEV to serve the *j*th flight that is awaiting service at Stage *t*. Therefore, $w_{jt}$ is a random discrete variable.

Now the continuous random variable $\widetilde{w}_{jt}$ is discretized into $w_{jt}$: first, divide the time range of $[b, a]$ into m stages at an interval of $\Delta t = 5$ minutes (assuming that the length of $[b, a]$ is $m\Delta t$). These m stages are represented by m integers from $b/\Delta t$ to $a/\Delta t - 1$, therefore, $w_{jt} \in [b/\Delta t, a/\Delta t - 1]$ and $w_{jt}$ is an integer. Secondly, the probability of $w_{jt}$ taking value *k* out of the m values is given by

$$\text{Prob}(w_{jt} = k) = \Phi(\mu, \sigma^2; \rho_u) - \Phi(\mu, \sigma^2; \rho_l) \quad (2)$$

where $\Phi$ is the cumulative distribution function as defined in (1). $\rho_u$ and $\rho_l$ are the upper and lower bounds of $\widetilde{w}_{jt}$ within Stage *k*, respectively. $w_{jt}$ is the discretized random workload, as explained above. For example, $w_{jt} \in \{3, 4, 5, 6\}$, meaning that the ground transport for the jth flight at Stage t requires 15 minutes ($w_{jt} = 3$ stages) to 30 minutes ($w_{jt} = 6$ stages) to complete. The probability of $w_{jt}$ taking each discrete value is given by (2). $w_{jt}$ is a critical input for the ASEV dynamics model introduced in Section 3.2.

**3.2 ASEV Dynamics Model**

In this section, an ASEV dynamics model is presented, which models the control decisions, the ASEV states and their transitions over time. The model considers the uncertain ground transport workload as modelled in Section 3.1.

At any Stage *t* (time is discretized into stages), the ASEV fleet state $S_t$ consists of the states of all individual ASEVs. $S_{it}$ denotes the state of an *i*th ASEV at Stage *t*, given by

$$S_{it} = [q_{it}, SoC_{it}, f_{Rit}] \qquad (3)$$

where $q_{it}$ is a discrete state: $q_{it} = 1$ means that the *i*th ASEV is charging at Stage *t*; $q_{it} = 0$ means that it is idling; $q_{it} < 0$ means that it is working (in this paper, "working" means undertaking ground transport) and it will take $|q_{it}|$ stages to complete the work. $SoC_{it}$, a continuous state, is the state of charge (SoC) of the *i*th ASEV's battery at Stage *t*. $f_{Rit}$ denotes the battery cycles to failure for the *i*th ASEV at Stage *t*.

Fig. 2 presents an overview of the ASEV dynamics model.

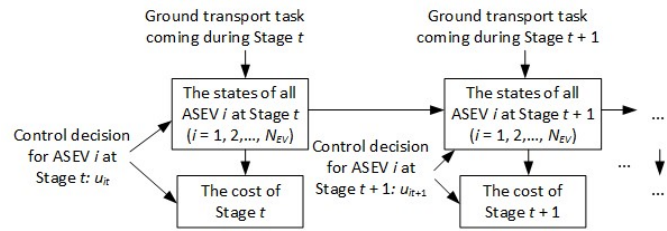

Fig. 2. Overview of the ASEV dynamics model

According to Fig 2, the optimal control is performed online, i.e. the control decision for each Stage *t* is made when the state $S_{it}$ at Stage *t* becomes known.

The energy cost for the $i$th ASEV at Stage $t$ is given by

$$C_{it} = \begin{cases} C_R \cdot max\{q_{it}, 0\} \cdot E_c & if\, max\{q_{it}, 0\} \cdot E_c \leq E_{Rt} \\ C_R E_{Rt} + C_{Gt}(max\{q_{it}, 0\} \cdot E_c - E_{Rt}) & otherwise \end{cases} \quad (4)$$

where $C_R$ denotes the energy price per kWh from renewable generation. $E_c$ denotes the energy consumption (some of the energy is charged to the battery and the rest is loss) during each stage. $E_c$ is a constant given the assumption of the constant battery charging power. $C_{Gt}$ denotes the price per kWh of the grid-supplied energy at Stage $t$. $E_{Rt}$ denotes the available energy generated by renewable generation at Stage $t$. $q_{it}$ is given in (3). The "max" term in (4) ensures that the energy cost is incurred only when the ASEV is charging. Equation (4) is based on the principle that the ASEV fleet give priority to consuming the cheap energy directly purchased from renewable generation over consuming the grid-supplied energy.

The battery degradation cost for the $i$th ASEV at Stage $t$ is given by

$$B_{it} = f(SoC_{it}, E_w, f_{Rit}) \quad if\ q_{it} < 0 \quad (5)$$

where $q_{it}$, $SoC_{it}$ and $f_{Rit}$ are given in (3). $E_w$ is the energy discharged during Stage $t$. Function $f$ is the linear function for battery degradation cost, with its coefficient derived from [23]. It is the function of the SoC, energy discharged during Stage $t$, and the cycles to failure.

The total cost (including energy and battery degradation costs) for all ASEVs at Stage $t$ is given by

$$g_t = \sum_{i=1}^{N_{EV}} (C_{it} + B_{it}) \quad t = 0,1,2\ldots, N-1 \quad (6)$$

where $C_{it}$ and $B_{it}$ are given in (4) and (5), respectively. $N_{EV}$ is the total number of ASEVs.

Ideally, all ASEV batteries should be charged to full at the end of the day to prepare the ASEVs for ground transport the next day. If any battery is not charged to full at the last stage of the day (Stage $N$), this incurs a terminal stage cost, as given by

$$g_N = \sum_{i=1}^{N_{EV}} C_{GN} B(SoC_{max} - SoC_{iN}) \qquad (7)$$

where $C_{GN}$ denotes the price per kWh of the grid-supplied energy at Stage $N$. $B$ denotes the battery energy capacity. $SoC_{max}$ is the upper bound of the SoC. $N_{EV}$ is the total number of ASEVs. $SoC_{iN}$ denotes the SoC of the *i*th ASEV at Stage $N$.

The objective of the ASEV optimal control is to minimize the summation of $g_t$ over all stages of the day.

$$\text{Min } J = g_N + \sum_{t=0}^{N-1} g_t \qquad (8)$$

where $g_t$ and $g_N$ are given in (6) and (7), respectively.

When the SoC of the *i*th ASEV battery reaches either the upper bound or the lower bound, there are two state constraints:

Case 1: an ASEV *i* is prevented from switching to work because of a low SoC.

$$\text{If } q_{it} \geq 0 \text{ and } SoC_{it} \leq SoC_{min}, \text{ then } q_{it+1} = u_{it} \geq 0 \qquad (9)$$

where $q_{it}$, $q_{it+1}$, and $SoC_{it}$ are defined in (3). $SoC_{min}$ denotes the lower bound of the SoC. $u_{it}$ is the control decision for ASEV *i* at Stage *t*: $u_{it} = 1$ means "to charge battery"; $u_{it} = 0$ means "to idle"; and $u_{it} = -1$ means "to work (i.e. undertake ground transport)".

Case 2: an ASEV *i* is prevented from battery charging because its SoC has reached the upper bound.

$$\text{If } q_{it} \geq 0 \text{ and } SoC_{it} = SoC_{max}, \text{ then } q_{it+1} = u_{it} \neq 1 \qquad (10)$$

where $q_{it}$, $q_{it+1}$, and $SoC_{it}$ are defined in (3). $SoC_{max}$ is defined in (7); $u_{it}$ is defined in (9).

When the SoC of the ith ASEV battery is above the lower bound and that the ASEV is not currently working, a control-based state transition can occur. This is further divided into two

cases:

Case 1: the ASEV i is controlled to work.

$$\text{If } SoC_{min} < SoC_{it} \text{ and } q_{it} \geq 0 \text{ and } u_{it} = -1, \text{ then } q_{it+1} = -w_{jt} \quad (11)$$

where $SoC_{min}$ is defined in (9). $q_{it}$, $q_{it+1}$, and $SoC_{it}$ are defined in (3). $u_{it}$ is defined in (9). $w_{jt}$ denotes the number of stages (the amount of time) required for an ASEV to serve the jth flight that is awaiting service at Stage t, as explained in Section 3.1.

When a flight j is awaiting ground transport service at Stage t, it should be served as soon as there is at least one free ASEV.

$$\text{If } w_{jt} > 0 \text{ and } \exists i: q_{it} \geq 0 \text{ and } SoC_{it} > SoC_{min}, \text{ then } \exists i: q_{it+1} = -w_{jt} \text{ and } u_{it} = -1 \quad (12)$$

where all variables are defined the same as in (11).

If a flight j is awaiting service at Stage t but because no ASEV is available, the service for flight j is delayed to Stage t + 1. This translates to:

$$\text{if } w_{jt} > 0 \text{ and } \forall i: q_{it} < 0, \text{ then } w_{jt+1} = w_{jt} \text{ and } d_j \leftarrow d_j + 1 \quad t = 0,1,2\ldots,N-1 \quad (13)$$

where $w_{jt}$ is defined in Section 3.1. $q_{it}$ is defined in (3). $d_j$ denotes the stages of delay. It is initialized to zero.

A hard constraint exists that the stages of service delay for any flight should be no more than a threshold.

$$d_j \leq d_{thre} \quad (14)$$

where $d_{thre}$ is the threshold of delay; $d_j$ is defined in (13).

Case 2: the ASEV i is controlled to idle or charge.

$$\text{If } SoC_{it} < SoC_{max} \text{ and } q_{it} \geq 0 \text{ and } u_{it} \geq 0, \text{ then } q_{it+1} = u_{it} \tag{15}$$

where $SoC_{max}$ is defined in (7). $q_{it}$, $q_{it+1}$, and $SoC_{it}$ are defined in (3). $u_{it}$ is defined in (9).

When the $i$th ASEV is working, its work cannot be interrupted by any control decision. The ASEV will naturally complete the work. This is expressed as

$$\text{If } q_{it} \leq -2, \text{ then } q_{it+1} = q_{it} + 1 \text{ and } u_{it} = -1 \tag{16}$$

$$\text{If } q_{it} = -1, \text{ then } q_{it+1} = 0 \text{ and } u_{it} = 0 \tag{17}$$

where all variables are defined the same as in (12).

Fig. 3 presents a state transition graph describing the relation among $q_{it}$, $w_{jt}$, and $u_{it}$.

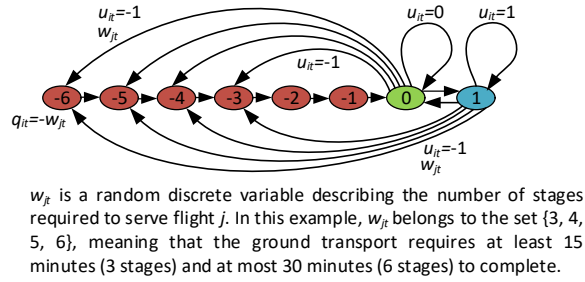

$w_{jt}$ is a random discrete variable describing the number of stages required to serve flight $j$. In this example, $w_{jt}$ belongs to the set {3, 4, 5, 6}, meaning that the ground transport requires at least 15 minutes (3 stages) and at most 30 minutes (6 stages) to complete.

Fig. 3. State transition graph for the $i$th ASEV.

In Fig. 3, each circle represents a state of the $i$th ASEV. The value in each circle is $q_{it}$, i.e. the discrete state of the $i$th ASEV at Stage $t$. Red circles mean that the $i$th ASEV is working. The green and blue circles mean that the $i$th ASEV is idling and charging, respectively. As mentioned above, work cannot be interrupted. Therefore, in Fig. 3, the state transits naturally from $-6$ to 0 over time, as described by (16) and (17).

For any ASEV $i$ at stage $t$, the continuous dynamics of its battery SoC depends on its control decision $u_{it}$.

$$SoC_{it+1} = \begin{cases} SoC_{it} - E_w & \text{if } u_{it} = -1 \\ SoC_{it} + \min\{\gamma E_c, SoC_{max} - SoC_{it}\} & \text{if } u_{it} = 1 \\ SoC_{it} & \text{if } u_{it} = 0 \end{cases} \quad (18)$$

where $SoC_{it}$ and $SoC_{it+1}$ are defined in (3). $u_{it}$ is defined in (9). $E_w$ is defined in (5). $E_c$ is the energy consumption during each stage, as defined in (4). $\gamma$ is the efficiency of the battery. $\gamma E_c$ is therefore the energy charged to the battery during each stage.

With the ASEV dynamics model established, the next step is to determine a sequence of control variables $u_{it}$ (defined in (9)) for all $i$ (all ASEVs) and for all $t$ (all stages of a day), with the objective to minimize the total operation cost $J$ (defined in (8)).

## 4. Near-optimal control of ASEV dynamics model

Based on the ASEV dynamics model detailed in the last section, a rollout approach is presented as a near-optimal control method to determine a sequence of control variables $u_{it}$ for each ASEV at each stage $t$.

At each stage $t$, the optimal cost-to-go function $J_t$ is defined as the minimum total cost from Stage $t$ to Stage $N-1$ (the last stage of the day) plus the terminal stage cost $g_N$ (as given by (7)). Because the prohibitively large number of states in the ASEV dynamics model cause a combinatorial explosion, it is impossible to calculate the accurate cost-to-go function $J_t$, thus being impossible to develop an accurate optimal control strategy for the ASEV dynamics model. A customized rollout approach is developed to yield a near-optimal control strategy through approximations. It consists of the following steps:

1) Two customized suboptimal heuristic control strategies are developed to approximate the cost-to-go function $J_m$ as $\tilde{J}_m$, given the starting state $S_m$ (the ASEV fleet state at Stage $m$). The two heuristics are elaborated as follows:

***Heuristic i)*:** the "renewable matching" heuristic. At each Stage *t* from Stage *m* to the last stage of the day, control the ASEVs to charge only when there is available renewable energy as dictated by the renewable generation profile. When a flight is awaiting ground transport service, always assign the available ASEV with the greatest SoC to take the work. The pseudo-code for heuristic i) is presented in Fig. 4.

***Heuristic ii)*:** the "greedy charging" heuristic. Given the starting state $S_m$ at Stage *m*, control the ASEVs to charge as early as possible until the maximum SoC is reached. When a flight is awaiting ground transport service, always let the available ASEV with the greatest SoC take the work. The pseudo-code for heuristic ii) is presented in Fig. 5.

Heuristic i) is not always feasible because, when renewable energy is seriously deficient throughout a day, the ASEV batteries all have too low SoC values to undertake "peak" workload of ground transport. If heuristic i) is not feasible from Stage *t*, then heuristic ii) is selected. If both heuristics are feasible from Stage *t*, the better one (the one that leads to a lower $\tilde{J}_t$) of the two heuristics is selected. The approximate cost-to-go $\tilde{J}_t$ for the selected heuristic is recorded for use in Step 2).

```
0. Initialize the approximate cost-to-go (starting from Stage m) as zero, i.e.
   J̃_m = 0.
1. Loop over all stages t from Stage m to the last stage of the day {
        2. Order all available (not currently working) ASEVs by increasing
           SoC values at Stage t.
        3. If a flight j is awaiting ground transport at Stage t, control the
           available ASEV i with the highest SoC value to serve this flight. Set
           this ASEV i as unavailable.
        4. Initialize the total energy consumption (the energy taken from the
           grid and renewable source by the ASEV fleet batteries) at Stage t to
           zero, i.e. E_t = 0.
        5. If the available renewable energy at Stage t is positive, i.e. E_Rt > 0 {
                6. Loop over all available ASEVs i from the one with the
                   lowest SoC value to the one with the highest SoC value {
                        7. Control ASEV i to charge if it is not currently
                           working and that it is not controlled to work.
                        8. The total energy consumption by the ASEV fleet
                           E_t ← E_t + E_c, where E_c is the energy charged to
                           ASEV i at Stage t.
                        9. If E_t ≥ E_Rt then exit the inner loop (E_Rt is the
                           available renewable energy at Stage t).
                }
        }
        10. Calculate g_t, the cost of Stage t, according to (6). Update the
            approximate cost-to-go: J̃_m ← J̃_m + g_t
}
11. Calculate g_N, the terminal stage cost, according to (7). Update the
    approximate cost-to-go: J̃_m ← J̃_m + g_N. Output the approximate cost-to-go J̃_m.
```

Fig. 4. Pseudo-code for heuristic i), i.e. the "renewable matching" heuristic.

```
0. Initialize the approximate cost-to-go (starting from Stage m) as
   zero, i.e. J̃_m = 0.
1. Loop over all stages t from the current Stage m to the last stage of
   the day {
        2. Order all available ASEVs by increasing SoC values at
           Stage t.
        3. If a flight j is awaiting ground transport at Stage t, control
           the available ASEV i with the highest SoC value to serve
           this flight. Set this ASEV i as unavailable.
        4. Initialize the total energy consumption (the energy taken
           from the grid and renewable source by the ASEV fleet
           batteries) at Stage t to zero, i.e. E_t = 0.
        5. Loop over all available ASEVs i {
                6. Control ASEV i to charge if it is not currently
                   working and that it is not controlled to work.
                7. The total energy consumption by the ASEV fleet
                   E_t ← E_t + E_c, where E_c is the energy charged to
                   ASEV i at Stage t.
        }
        8. Calculate g_t, the cost of Stage t, according to (6). Update
           the approximate cost-to-go: J̃_m ← J̃_m + g_t
}
9. Calculate g_N, the terminal stage cost, according to (7). Update the
   approximate cost-to-go: J̃_m ← J̃_m + g_N. Output the approximate
   cost-to-go J̃_m.
```

Fig. 5. Pseudo-code for heuristic ii), i.e. the "greedy charging" heuristic.

2) Given $S_t$ (the ASEV fleet state at Stage $t$) which consists of $S_{it}$ for all ASEVs $i$, the rollout approach generates the set of all possible $S_{t+1}$ by enumerating all feasible control decisions $u_{it}$ (defined in (9)) for Stage $t$, considering the workload $w_{jt}$ (defined in (11)). The approach then selects the "best" $S_{t+1}$ that produces the minimum approximate cost-to-go among all $S_{t+1}$ in the set [20]. The mathematical expression is

$$S_{t+1} = \mathrm{argmin}_{S \in N(S_t)} \ J(S) \tag{19}$$

Where $S_t$ is the state at Stage $t$. $N(S_t)$ is the set of all possible states at Stage $t+1$. $J(S)$ is the approximate cost-to-go $\tilde{J}_{t+1}$ of the better one of the two heuristics, expressed as the function of state $S$. The rollout control $u_{it}$ for all ASEVs $i$ is the control that corresponds to the transition from $S_t$ to $S_{t+1}$.

An alternative expression with the same meaning is given by

$$u_t = \mathrm{argmin}_{u_t \in U_t \text{ and } S \in N(S_t)} [g_t + J(S)] \tag{20}$$

where $u_t$ is the set of control decisions for all ASEVs $i$ at Stage $t$, i.e. $u_t = \{u_{it} \text{ for all } i\}$. $U_t$ is the constraint set for $u_t$ at Stage $t$. $S_t$, $N(S_t)$, and $J(S)$ are defined in (19). $g_t$ is defined in (6).

This process iterates until $S_t$, $S_{it}$, and $u_{it}$ for all stages $t$ are determined. The sequence of $u_{it}$ for all ASEVs $i$ and for all stages $t$ constitute a near-optimal control strategy, which controls each individual ASEV to charge, idle, and work at each stage.

## 5. Case Study

In this section, case studies are performed to validate the ASEV dynamics model and the customized rollout approach. The case studies are based on Bristol Airport, a medium-sized airport in the UK. A typical day's flight information is obtained from Bristol Airport website [24]. Considering the scale of the airport, the number of ASEVs is set as 25. The renewable power output profiles for summer and winter are obtained from [25] and [26], respectively. The battery charging type is fast charging at a constant power of 22 kW [27]. The battery cycle efficiency is 90% [28]. To prevent overcharge and deep discharge, the SoC of each battery is kept between 20% and 80%. The battery capacity is 50 kWh [29]. The case studies consider photovoltaic (PV) generation. The price for the PV energy is £0.04/kWh. The tariffs of the grid-supplied energy follow a two-tier tariff system, which is shown in Fig. 6.

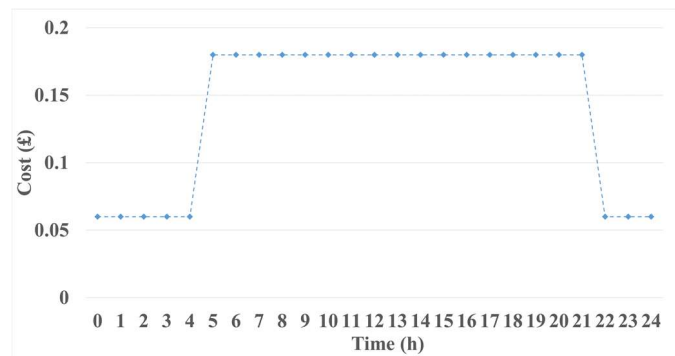

Fig. 6. The two-tier tariff for one day.

To validate the algorithm, this section performs three sets of simulations. The first set is a comparison among the two heuristic algorithms and the rollout algorithm on a typical summer day. The second set is a comparison among the two heuristic algorithms and the rollout algorithm on a typical winter day, which corresponds to a substantially different PV output profile from that on the summer day. The third set of simulations consider the cancellation of flights at short notice and evaluate the performance of the rollout algorithm under such circumstances.

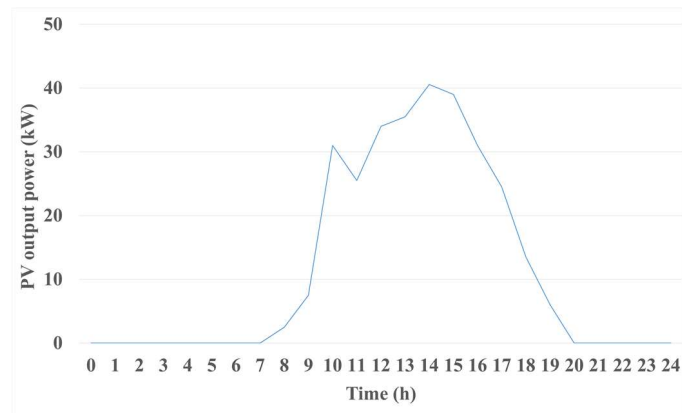

Fig. 7. The PV generation power of the typical sunny summer day.

**Scenario 1): Comparison between the heuristic algorithm and the rollout algorithm for a typical sunny day in summer**

A typical sunny day in summer is chosen for the case study. Fig. 7 shows the PV output profile for the day. On the day, there are 88 flights arriving at Bristol Airport and 86 flights departing. The earliest one is at 6.00 am and the last one is at 10.55 pm. The workload of serving any given flight is a random variable. The random workload model is explained in Section 3.1. Each flight is served by one ASEV. At the start of the day, the SoC of each ASEV is at the maximum value $SoC_{max}$ and all ASEVs are idling (i.e. ready to work).

The three control algorithms, Heuristic i), Heuristic ii), and the rollout algorithm, are compared with each other. The results are shown in Fig. 8.

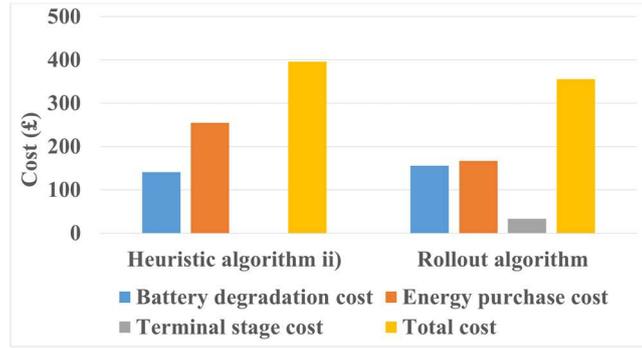

Fig. 8. The cost of Bristol Airport under Heuristic ii) and rollout algorithm on the sunny summer day.

When Heuristic i) is applied, no ASEV is available to serve any flight at stage 208 (5.15pm). It means that Heuristic i) is infeasible. This is because Heuristic i) only charges the ASEV batteries when there is PV energy available. If the PV energy is deficient throughout the day, Heuristic i) would charge the ASEVs insufficiently, whereas they are still controlled to work (i.e. serve the flights). In this case, all ASEV have too low SoC values to serve any flight at 5.15pm of the day and that at least one flight remains unserved for two consecutive stages, resulting in Heuristic i) being infeasible.

If Heuristic ii) is applied, the total cost is £395.71 for the airport on the day. The battery degradation cost, energy purchase cost and terminal stage cost are £140.71, £254.99, and £0, respectively.

If the rollout algorithm is applied, the total cost is £357.95 for the airport on the day. This is broken down into the battery degradation cost, energy purchase cost, and terminal stage cost of £156.32, £168.56, and £33.08, respectively.

Table I shows the control decisions for an example ASEV on a typical summer day under the rollout algorithm.

TABLE I

CONTROL STRATEGY OF AN ASEV UNDER ROLLOUT ALGORITHM ON A SUNNY SUMMER DAY

| Charging | | Working | |
|---|---|---|---|
| Start time | End time | Start time | End time |
| 08.00 | 08.15 | 07.35 | 07.45 |
| 10.40 | 11.10 | 10.10 | 10.30 |
| 12.55 | 13.25 | 12.30 | 12.45 |
| 14.10 | 14.50 | 13.45 | 14.00 |
| 15.25 | 17.10 | 14.55 | 15.15 |
| 17.55 | 20.00 | 17.25 | 17.45 |
|  |  | 20.55 | 21.10 |

*The other time slots not shown in the table correspond to the ASEV idling

From the comparison of the two heuristics and the rollout algorithm, it is clear that the rollout algorithm incurs a total cost 10.5% less than the cost of Heuristic ii). Heuristic i) is infeasible. The battery degradation cost and energy purchase cost of the rollout algorithm are 11.1% more than and 51.3% less than those of Heuristic ii), respectively. The rollout algorithm achieves a significant saving of the energy purchase cost, compared to Heuristic ii), because: Heuristic ii) does not care about the electricity price at all but charges the battery whenever the SoC is not the maximum. In contrast, the rollout algorithm takes advantage of both the cheap PV energy and the low tariff period of the grid-supplied energy.

The battery degradation cost under Heuristic ii) is less than that under the rollout algorithm. This is because Heuristic ii) performs a greedy charging, with its average SoC at the start of charging being greater than that under the rollout algorithm. In other words, the rollout algorithm leads to deeper discharges and thus a greater battery degradation cost than Heuristic ii). The terminal stage cost of Heuristic ii) is £0 because Heuristic ii) charges an ASEV battery whenever it is not full and that the ASEV is not working, regardless of the electricity price. This ensures that the ASEV batteries all have the maximum SoC value at the end of the day, resulting in a zero terminal stage cost. In contrast, the rollout algorithm only charges the ASEV batteries when the electricity price is cheap. As a result, at the end of the day not all ASEV batteries are fully charged, causing a positive terminal stage cost.

**Scenario 2): Comparison between Heuristic ii) and the Rollout algorithm for a typical sunny day in winter**

Scenario 1 is a typical sunny day in summer in the UK. Scenario 2 considers a typical sunny day in winter in the UK. The flight schedule is the same as in Scenario 1). However, the daytime is significantly different, resulting in a different PV generation profile. Fig. 9 shows the output profile of the PV generation on the sunny day in winter [21]. The price curve of grid-supplied energy is the same as that of Scenario 1). The starting state of the ASEVs is also the same as in Scenario 1). The results are shown in Fig. 10.

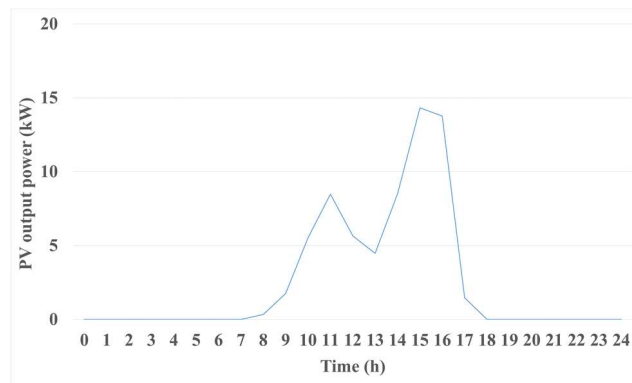

Fig. 9. The PV generation power of the typical sunny winter day.

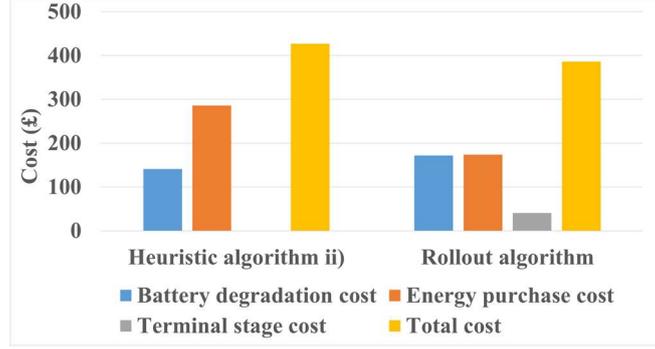

Fig. 10. The cost of Bristol Airport under Heuristic ii) and rollout algorithm on the winter sunny day.

Table II presents the control decisions for the example ASEV on a typical winter day under the rollout algorithm.

TABLE II

CONTROL STRATEGY OF AN ASEV UNDER ROLLOUT ALGORITHM ON A SUNNY WINTER DAY

| Charging | | Working | |
|---|---|---|---|
| Start time | End time | Start time | End time |
| 07.10 | 07.40 | 06.40 | 07.00 |
| 09.35 | 09.55 | 09.10 | 09.25 |
| 12.45 | 13.15 | 12.15 | 12.35 |
| 14.10 | 14.40 | 13.40 | 14.00 |
| | | 16.15 | 16.35 |
| | | 20.55 | 21.15 |

*The other time slots not shown in the table correspond to the ASEV idling

The total costs under Heuristic ii) and the rollout algorithm are £426.67 and £386.33, respectively. For Heuristic ii), the battery degradation cost is £140.71, the energy purchase cost is £285.95 and the terminal stage cost is £0. For rollout algorithm, the battery degradation cost is £171.97, the energy purchase cost is £173.61 and the terminal stage cost is £40.76. The above results demonstrate a similar trend to the results in Scenario 1). For the rollout algorithm, the battery degradation cost is greater than that under Heuristic ii) by 18.2%, whereas the energy purchase cost is 64.4% less.

Despite the difference in the PV output profile between winter and summer, the rollout algorithm yields satisfactory results and a lower total cost than Heuristic ii) in both scenarios. This demonstrates the general applicability of the rollout algorithm for different seasons.

**Scenario 3): The case where there are flights cancelled at short notice**

The rollout algorithm runs on a receding time horizon. This section considers the case with flight cancellations at short notice. Both the rollout algorithm and Heuristic ii) are applied and their results are compared with each other. Suppose that the arrival flight at 10.45am is cancelled. Other conditions remain the same as mentioned in Scenario 1). The comparison results are shown in Fig. 11.

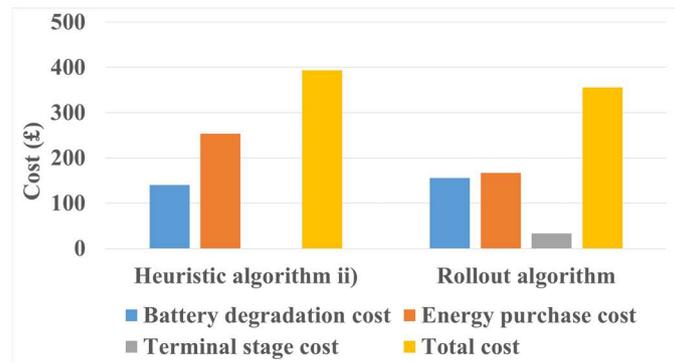

Fig. 11. The cost of Bristol Airport under Heuristic ii) and rollout algorithm when flights are cancelled at 10.45am.

Table III shows the control decision for an example ASEV under the rollout algorithm. Under the rollout algorithm, the battery degradation cost is £155.62, the energy cost is £167.07, the terminal stage cost is £33.08, and the total cost is £355.77. Under Heuristic ii), the total cost is £395.71. The battery degradation cost is £140.71, the energy purchase cost is £254.99 and the terminal stage cost is £0.

TABLE III

CONTROL STRATEGY OF AN EXAMPLE ASEV UNDER ROLLOUT ALGORITHM WHEN A FLIGHT IS CANCELLED AT 10.45AM.

| Charging | | Working | |
| --- | --- | --- | --- |
| Start time | End time | Start time | End time |
| 06.45 | 07.05 | 06.20 | 06.35 |
| 09.05 | 09.35 | 08.35 | 08.55 |
| 12.30 | 12.50 | 12.05 | 12.20 |
| 13.40 | 14.10 | 13.10 | 13.30 |
| 16.20 | 16.50 | 15.50 | 16.10 |
| | | 17.30 | 17.50 |
| | | 21.05 | 21.20 |

*The other time slots not shown in the table correspond to the ASEV idling

The case study proves that both the rollout algorithm and Heuristic ii) are adaptive towards the changing flight schedule. But the total cost under the rollout algorithm is 10.6% less than that under the Heuristic ii). The comparisons of the three cost components (i.e. the battery degradation cost, the energy purchase cost and the terminal stage cost) show a similar trend to the comparisons in Scenario 1). The rollout algorithm yields a battery degradation cost 11.1% greater than that under Heuristic ii) and an energy purchase cost 51.7% less than that under Heuristic ii). Heuristic ii) leads to a zero terminal stage cost, whereas the rollout algorithm leads to a terminal stage cost of £33.08. The reason for having such a trend is the same as that explained in Scenario 2).

From the simulation results, the rollout algorithm is better than the Heuristic ii). It is adaptive towards the changes of the environment in real time, e.g. flight cancellations at short notice.

## 6. Conclusions

This paper proposes a new dynamics model for airport service electric vehicles (ASEVs) and a new customized rollout approach as a near-optimal control method for the ASEV dynamics model. Case studies compare the rollout algorithm and the two heuristics (one heuristic matches the renewable generation and the other heuristic charges the battery whenever its SoC is not the maximum) for a typical sunny summer day and a typical sunny winter day. The two days have very different PV output profiles. In both cases, the rollout algorithm achieves a lower total cost than Heuristic ii). This is because the rollout algorithm takes advantage of the cheap PV energy as well as the off-peak price of the grid-supplied energy. However, Heuristic i) is infeasible in both cases, because the PV energy is insufficient to support the ASEVs for a whole day's work. The case study also demonstrated that the rollout algorithm is adaptive towards flight schedule changes. When there is a flight cancellation at short notice, the rollout algorithm achieves a lower total cost than the Heuristic ii). The research outcome will save costs and reduce carbon emissions for the airport in the context of transport electrification, facilitate the local consumption as well as the penetration of distributed generation, and make the battery-charging load friendly to the grid.